\documentclass[12pt]{article}
\usepackage{psfig}

\newcommand{\beq}{\begin{equation}}
\newcommand{\eeq}{\end{equation}}
\newcommand{\bea}{\begin{eqnarray}}
\newcommand{\eea}{\end{eqnarray}}

\begin{document}

\title{{\bf Low $Q^2$ wave-functions of pions and kaons and their parton 
distribution functions.}}
\author{{\bf C. Avila}\thanks{e-mail: cavila@uniandes.edu.co}~, ~  
{\bf J. Magnin}\thanks{e-mail: jmagnin@uniandes.edu.co} ~ and 
{\bf J.C. Sanabria}\thanks{e-mail: jsanabri@uniandes.edu.co}\\
{\normalsize {\it Depto. de F\'{\i}sica, Universidad de los Andes,}}
\\ {\normalsize {\it AA 4976, Santaf\'e de Bogot\'a, Colombia }}
}

\maketitle
\date

\begin{abstract}
We study the low $Q^2$ wave-functions of pions and kaons as  
an expansion in terms of hadron-like Fock state fluctuations. 
In this formalism, pion and kaon wave-functions are related one another.
Consequently, the knowledge of the pion structure allows the determination 
of parton distributions in kaons. In addition, we show that the intrinsic (low $Q^2$) 
sea of pions and kaons are different due to their different valence 
quark structure. Finally, we analize the feasibility of a method to extract 
kaon's parton distribution functions within this approach and 
compare with available experimental data.
\end{abstract}

\section{Motivation}

At present the only information available on the structure of unstable mesons is 
on parton distributions functions (pdf) in pions. These pdf are extracted 
from Drell-Yan (D-Y) dilepton and prompt photon production in pion-nucleon interactions. 

In principle, the same type of experiments can be used to extract information on 
parton distributions in kaons by only replacing the pion beam by a kaon one. However, 
this requires high intensity kaon beams which are not easily attainable in present day 
experiments. Furthermore, there is an additional difficulty which is inherent to the kaon 
structure because strange valence quarks in the kaon must anihilate with strange quarks in the 
target particles. As targets are made of nucleons, and strange 
quarks are in the sea of them, these $s-\bar{s}$ anihilation processes are  expected to have a small 
contribution to the total D-Y dilepton cross section. Consequently, it is very difficult to 
have a precise measurement of the strange valence quark distribution in kaons.
Other procedures involving the detection of a D-Y dilepton pair accompanied by a fast pion in 
the final state have been proposed to measure the strange quark distribution in kaons~\cite{rusos}, 
but they rely on several assumptions coming from recombination models.

The situation, however, is radically different with the light valence quark distribution. In 
fact, the $\bar{u}_K$ distribution can be measured rather well in D-Y experiments since this is, 
by far, the major contribution to $q\bar{q}\rightarrow l^+l^-$ processes in $K^--nucleon$ interactions. 
The first, and up to our knowledge, the only attempt in this direction was made by the NA3 
Collaboration~\cite{na3}, who measured the ratio of the $u$-quark distribution in kaons to the 
$u$-quark distribution in pions, $\bar{u}_K/\bar{u}_\pi$. 
In this experiment, the ratio $\bar{u}_K/\bar{u}_\pi$ was extracted from D-Y dimuon 
production in $150$ GeV/$c$ $K^-,\pi^--nucleus$ 
interactions. The measurement is, however, subject to large uncertainties due to the 
limited statistic of the experiment and several assumptions made in the calculation of 
the $\bar{u}_K/\bar{u}_\pi$ ratio.

Another possibility is to extract information on 
the kaon structure from precise measurements of the pion's pdf. 
To achieve this end, a model of the low $Q^2$ wave-function of hadrons in terms of 
hadron-like Fock state fluctuations is called for. In this model, as we will show, the pion 
and kaon low $Q^2$ wave-functions are related to each other. This allows in principle to 
extract information on the low $Q^2$ structure of kaons from the 
pion structure by means of a rather direct procedure. 
It is worth to note that this procedure 
only needs the measurement of the pion's pdf and the ratio $\bar{u}_K/\bar{u}_\pi$, 
avoiding in this way any measurement of the strange distribution in kaons, which is very 
complicated from the experimental point of view.

In Refs.~\cite{christiansen,magnin,christiansen2}, a model of nucleons in terms of 
meson-baryon bound state fluctuations has been developed. This model 
describes well the $\bar{d}-\bar{u}$ and the $\bar{d}/\bar{u}$ asymmetries in the 
nucleon sea as measured by the E866 Collaboration~\cite{e866}. The model also describes qualitatively 
the $s-\bar{s}$ asymmetry in nucleons observed in a recent global analysis of Deep Inelastic 
Scattering (DIS) data in nucleons~\cite{barone}. Moreover, it is interesting to note that this model 
provides a consistent scheme to generate non-perturbative sea quark and gluon distributions 
at a low $Q^2$ scale. Let us remark that, as already noted by several authors~\cite{non-p}, 
these non-perturbative, valence-like, sea quark and gluon distributions at the low $Q^2$ input scale 
are needed to fit DIS data.

The above facts indicate that this could be a sensible approach to the problem of having a low $Q^2$ 
model of hadrons. In this work, we will extend the model to pions and kaons and we shall 
explore the feasibility of extracting information on the low $Q^2$ structure of kaons within this 
scheme.

The paper is organized as follows. In the next section we will study the low $Q^2$ wave-function 
of pions and kaons. In section~\ref{sec3}, we shall try to extract information on the kaon's pdf 
in terms of the model and the available experimental data and 
section~\ref{sec4} is devoted to conclusions and further discussion.

\section{Low $Q^2$ wave-functions of pions and kaons}
\label{sec2}

To start with, let us consider the $\pi^-$ and $K^-$. Their low $Q^2$ wave-functions 
can be expanded as 
\bea
\left|\pi^-\right> &=& a_0^{\pi}\left|\pi^-_0\right> + a_1^{\pi}\left|\pi^-g\right> 
+ a_2^{\pi}\left|K^\circ K^-\right> \nonumber \\
\left|K^-\right> &=& a_0^K\left|K^-_0\right> + a_1^K\left|K^-g\right> 
+ a_2^K\left|{\overline K}\,^\circ \pi^-\right> \; ,
\label{eq1}
\eea
in terms of hadron-like Fock state fluctuations. 
In the above expresions we have neglected higher order contributions involving 
heavier mesons. These fluctuations should be far off-shell and 
they can be safely ignored at this point. Coefficients in the expansions of eqs.~(\ref{eq1}) are 
constrained by probability conservation, $\sum_i{(a_i^M)^2}=1$.

Similar expresions can be obtained straighforwardly for $\pi^+$, $K^+$ and neutral kaons. 
Neutral pions, however, have a different structure than the charged ones because they can also 
fluctuate to $\left|\pi^+\pi^-\right>$ Fock states (see discussion in Ref.~\cite{christiansen2}).

Hadrons in the Fock state fluctuations on the RHS of eqs.~(\ref{eq1}) are assumed 
to be formed only by constituent valence quarks or {\em valons}, in the terminology of Ref.~\cite{hwa}. 
Fluctuations themselves are responsible for the non-perturbative -{\em intrinsic}- sea, which should 
provide the necessary binding among constituent quarks to form hadrons~\cite{christiansen2,brodsky}. 
This means that, at a low $Q^2$ scale, hadrons are made of constituent valons 
plus intrinsic quark-antiquark pairs and gluons. At this point it is important to note that valence 
quark densities at the low $Q^2$ scale have contributions not only from valons in the bare 
meson state, but also 
from the fluctuations themselves. These fluctuations give rise also to the so called intrinsic 
sea of $q\bar{q}$ pairs and gluons.

DGLAP evolution to higher $Q^2$ 
generates the perturbative -{\em extrinsic}- sea of 
$q\bar{q}$ pairs and gluons. In this sense, evolution to higher $Q^2$ should mostly resolve the 
structure of valons themselves~\cite{hwa}.

Let us call $Q_v^2$ the scale at which the hadron 
wave-function can be consistently written as a hadron-like Fock state expansion as 
those of eq.~(\ref{eq1}). Then, at the scale $Q_v^2$, parton distribution functions are given by
\bea
\bar{u}_{\pi}(x) &=& d_{\pi}(x) = (a_0^\pi)^2 v_{\pi}(x) + 
(a_1^\pi)^2 P_{\pi g} \otimes  v_{\pi} + (a_2^\pi)^2 P_{KK} \otimes v_{K} \nonumber \\
s_\pi(x) &=& \bar{s}_\pi(x)= (a_2^\pi)^2 P_{KK} \otimes v_{s/K} \nonumber \\
g_\pi(x) &=& (a_1^\pi)^2 P_{g\pi}(x)
\label{eq2}
\eea
for pions, and
\bea
\bar{u}_K(x) &=& (a_0^K)^2 v_{\bar{u}/K}(x) + 
(a_1^K)^2 P_{Kg} \otimes  v_{K} + (a_2^K)^2 P_{\pi K} \otimes v_{\pi} \nonumber \\
s_K(x) &=& (a_0^K)^2 v_{s/K}(x) + 
(a_1^K)^2 P_{Kg} \otimes  v_{s/K} + (a_2^K)^2 P_{K\pi} \otimes v_{s/K} \nonumber \\
d_K(x) &=& (a_2^K)^2 P_{\pi K} \otimes v_{\pi} \nonumber \\
\bar{d}_K(x) &=& (a_2^K)^2 P_{K\pi} \otimes v_{K} \nonumber \\
g_K(x) &=& (a_1^K)^2 P_{gK}(x)
\label{eq3}
\eea
for kaons. In eqs.~(\ref{eq2}) and (\ref{eq3})
\beq
P_{MM'} \otimes v_{q/M} \equiv \int_x^1{\frac{dy}{y}P_{MM'}(y)
v_{q/M}\left(\frac{x}{y}\right)}
\label{eq4}
\eeq
is the probability density of the non-perturbative contribution to the parton distribution, 
coming from the $\left|MM'\right>$ fluctuation~\cite{magnin,christiansen2}. $v_{q/M}$ is the 
$q$-flavored valon distribution in the meson $M$. A brief description about how to obtain the 
$\left|P_{MM'}\right>$ probability densities is given in the Appendix. In order to arrive to 
eqs.~(\ref{eq2}) and (\ref{eq3}), we have assumed 
\bea
v_\pi(x) &\equiv & v_{\bar{u}/\pi^-}(x) = v_{d/\pi^-}(x) = v_{u/\pi^+}(x) 
= v_{\bar{d}/\pi^+}(x) \nonumber \\
v_K(x) &\equiv & v_{\bar{u}/K^-}(x) = v_{d/K^0}(x)= v_{\bar{d}/{\overline K}^\circ}(x)
= v_{u/K^+}(x) \nonumber \\
v_{s/K} &\equiv & v_{s/K^-}(x) = v_{\bar{s}/K^+}(x) = v_{s/K^\circ}(x) = 
v_{\bar{s}/\bar{K}^\circ}(x)
\label{eq4a}
\eea
which follow from isospin invariance. 

Similar expresions to those of eqs.~(\ref{eq2}) and (\ref{eq3}) can be easily obtained for the 
$\pi^+$ and $K^+$ parton distribution functions.

It must be noted that there are no fluctuations containing neutral pions in the 
expansions of eq.~(\ref{eq1}). In fact, as discussed in Ref.~\cite{christiansen2}, 
pions cannot fluctuate to a $\left|\pi^{\pm,\circ}\pi^\circ\right>$ Fock state. This would 
imply the formation of a neutral in-pion state which can rapidly decay to a gluon. Actually, 
an unflavored structure like $v_q\,\bar{q}$ has the quantum numbers of a gluon, giving 
rise to the $\left|\pi^{\pm,\circ} g\right>$ Fock state. Consequently, the first contribution 
to the intrinsic $q\bar{q}$ sea in pions arises in the strange sector. Intrinsic  $u\bar{u}$ 
and $d\bar{d}$ sea quarks, however, must be considered in 
neutral pions due to contributions coming from the $\left|\pi^+\pi^-\right>$ 
fluctuation~\cite{christiansen2}.

Similarly, kaons cannot fluctuate to $\left|K^{\pm,\circ}\pi^\circ\right>$ bound states, which 
in turn are the origin of the $\left|K^{\pm,\circ}g\right>$ fluctuation. Consequently, in kaons, the 
first contribution to the intrinsic $q\bar{q}$ sea arises in the light sector. In fact, the 
$\left|{\overline K}^\circ\pi^-\right>$ ($\left|K^\circ\pi^+\right>$) fluctuation gives rise 
to the $d\bar{d}$ intrinsic sea of the $K^-$ ($K^+$) while the 
$\left|K^+\pi^-\right>$ ($\left|K^-\pi^+\right>$) one originates the $u\bar{u}$ intrinsic 
sea in the ${\overline K}^\circ$ ($K^\circ$) respectively. 
It is interesting to note also that there is no $s\bar{s}$ intrinsic 
sea in kaons. The reason is that fluctuations to $\left|K^{\pm,\circ}\phi\right>$, which could 
potentially originate the strange intrinsic sea, also contribute to the intrinsic gluon distribution 
due to the rapid anihilation of the unflavored $v_{s/K}\bar{s}$ structure. However, this 
contribution must be smaller than those coming from the anihilation of a light valon with a light 
antiquark. 

A careful look at the eqs.~(\ref{eq2}) and (\ref{eq3}) reveals other interesting properties of the 
low $Q^2$ wave-functions of pions and kaons. First of all, if the $v_\pi$, $v_K$ and 
$v_{s/K}$ valon distributions are known, the entire structure 
of pions and kaons is determined up to the coefficients giving the probability of each 
individual fluctuation. A second interesting property is that the intrinsic strange 
sea in pions is entirely determined by the strange valon distribution in kaons. 
Then, by using the convolution theorem applied to Mellin transforms we can formally invert the 
second expression in eq.~(\ref{eq2}) to obtain $v_{s/K}$:
\beq
v_{s/K}(n)=\frac{s_\pi(n)}{(a_2^\pi)^2P_{KK}(n)}\; ,
\label{eq10a}
\eeq
where $f(n)$ is the Mellin transform of $f(x)$. Notice that the above equation not only 
determines the shape of $v_{s/K}$, but also fixes the coefficient $a_2^K$, since the strange 
valon distribution must be normalized to one.

Once $v_{s/K}(x)$ is known, the $v_K(x)$ distribution become fixed by momentum 
conservation, thus
\beq
v_K(x) = v_{s/K}(1-x)\; .
\label{eq11}
\eeq

Then, by only measuring the strange quark distribution in pions, the valon probability densities 
in kaons become determined. Now, by measuring the $\bar{u}_\pi$ and $g_\pi$ we can determine the remaining 
$v_\pi$ through
\beq
v_\pi(n) = \frac{\bar{u}_\pi(n)-(a_2^\pi)^2 P_{KK}(n)\, v_K(x)}
{(a_0^\pi)^2+(a_1^\pi)^2 P_{\pi g}(n)}\; ;
\label{eq12}
\eeq
where we have used again the convolution theorem applied to Mellin transforms. 
Since the coefficients $a_1^\pi$ and $a_2^\pi$ are fixed once the strange and gluon 
distributions in pions are measured, then there are no free parameters in eq.~(\ref{eq12}) 
due to probability conservation.

To extract the shape of the required valon distributions and the value of the coefficients 
$a_i^\pi$ ($i=1,2,3$), data on $\pi^\pm-nucleon$ D-Y dilepton cross sections must be fitted using 
the formulae presented in eqs.~(\ref{eq2}). On this respect, it is worth to note that 
$\pi p \rightarrow \gamma X$ cross sections are dominated by $qg$ scattering, then they 
are sensitive to the gluon distribution. On the other hand, the difference of cross sections 
$\sigma(\pi^- p)-\sigma(\pi^+ p)$ is dominated by $q\bar{q}$ anihilation, allowing in this 
way the determination of quark distributions in pions. Furthermore, in Ref.~\cite{londergan} 
it has been argued that adequate linear combinations of $\pi^\pm-nucleon$ D-Y cross sections 
should allow the determination of valence densities in pions, independently of sea 
quark distributions. Having 
the gluon and valence quark distributions, the sea quark densities can be known. Thus, the 
determination of the valon densities $v_\pi$, $v_K$ and $v_{s/K}$ and the parameters $a_i^\pi$ 
can be achieved.

In practice, however, some sort of iterative procedure is needed to determine $v_\pi$, $v_K$ and 
$v_{s/K}$ from experimental data since $v_\pi$ enters in the calculation of the 
$P_{\pi g}$ and $P_{KK}$ probability densities.

At this point some remarks are in order. As we have shown above, the valon distributions and 
parameters entering in the pion wave-function can be determined from measurements of the pion pdf. 
Having this information, we can proceed to determine the kaon pdf. To this end, however, some 
experimental input on kaon parton distributions is needed. The $\bar{u}_K$ distribution can be 
measured in $K^--nucleon$ D-Y experiments, then we can use the relationship in the first line 
of eqs.~(\ref{eq3}) 
to fix the coefficients $a_i^K$ ($i=1,2,3$) in the kaon wave-function. Since $v_K$ and $v_{s/K}$ 
have been previously determined, then the kaon pdf become completely fixed. This will be done in 
the next section.

\section{Kaon parton distributions}
\label{sec3}

The existing parametrizations of pion pdf have been obtained 
assuming an $SU(3)_{flavor}$ symmetric sea. Furthermore, in most cases the $q\bar{q}$ sea has 
been generated perturbatively through DGLAP evolution~\cite{varios,grv-p}. Then it is not possible 
to determine the valon distributions $v_\pi$, $v_K$ and $v_{s/K}$ 
from present parametrizations. This will require to reanalyse the 
existing data on $\pi^\pm-nucleon$ dilepton and prompt  photon production in terms of the low $Q_v^2$ 
parametrizations of eqs.~(\ref{eq2}). Consequently, for the moment we refrain from any attempt 
of determining 
the valon distributions from experimental data. Instead, we will use the valon distribution 
in pions and kaons proposed in Ref.~\cite{hwa}, 
\bea
v_\pi(x) &=& 1 \nonumber \\
v_K(x)   &=& \alpha_K x^{a-1}(1-x)^{b-1} \nonumber \\
v_{s/K}(x) &=& \alpha_K x^{b-1}(1-x)^{a-1} \; ,
\label{eq14}
\eea
where $a$ and $b$ are related by $a/b=m_l/m_h\sim2/3$, $m_l$ and $m_h$ are the light and strange 
valon masses in kaons and $\alpha_K$ is a normalization constant.

These simple valon parametrizations would allow us to have both an idea of the feasibility 
of this method to extract kaon's pdf, as well as to have some knowledge 
of the kaon low $Q^2_v$ structure.

To obtain the values of the $a_i^K$ coefficients in the kaon wave-function, data on 
$\bar{u}_k/\bar{u}_\pi$ by the NA3 Collaboration~\cite{na3} were fitted using
\beq
\frac{\bar{u}_K(x)}{\bar{u}_\pi(x)} = 
\frac{(a_0^K)^2\,v_K(x) + (a_1^K)^2P_{Kg} \otimes v_K(x)+(a_2^K)^2P_{\pi K}\otimes v_\pi(x)}
{\bar{u}_\pi(x)}\; ,
\label{eq13}
\eeq
together with the valon distributions of eqs.~(\ref{eq14}). Notice that the fitting 
function has only two free parameters due to probability conservation.

In order to do the fits we assumed that the $\bar{u}_K/\bar{u}_\pi$ ratio is independent of $Q^2$. 
We recognize that this assumption could be questionable, but given the large error bars of 
the NA3 data, and the uncertainty in the valons distributions, fits are hardly sensitive to QCD evolution.

In these fits we also used two different $\bar{u}_\pi$ distributions; the GRV-P~\cite{grv-p} 
parametrization at the low $Q_v^2$ and the $\bar{u}_\pi$ distribution obtained by means of 
a Monte Carlo model of hadrons in Ref.~\cite{suecos}. This enables us to test the sensibility 
of the model to the shape of the $\bar{u}_\pi$ distribution.

\begin{figure}[b] 
\centerline{\psfig{figure=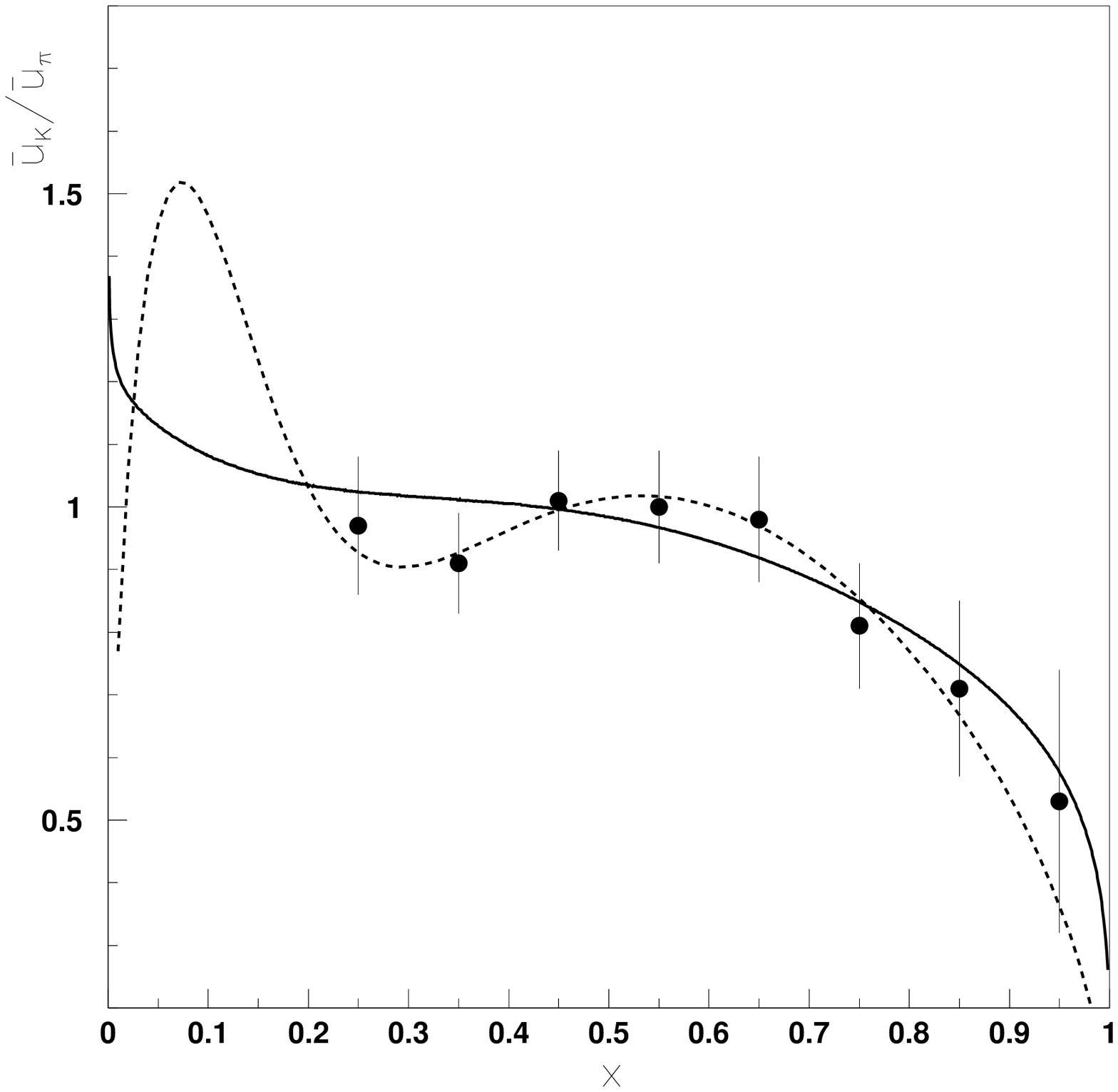,height=6.0in}}
\caption {$\bar{u}_K/\bar{u}_\pi$ as a function of x. Data is from Ref.~\cite{na3}. Dashed line 
is the fit using the $\bar{u}_\pi$ distribution of Ref.~\cite{grv-p}, the solid line is the fit using 
the $\bar{u}_\pi$ distribution given in Ref.~\cite{suecos}.}
\label{fig1}
\end{figure}

The results of the fits are displayed in Fig.~(\ref{fig1}) and in Table~(\ref{table1}). 
All the fits were done using $a=1.5$ and $b=2.25$ in the $v_K$ distribution of eqs.~(\ref{eq14}).
\begin{table}[t]
\begin{center}
\caption{Parameters in the $\bar{u}_K/\bar{u}_\pi$ fitting function.}
\label{table1}
\begin{tabular}{c|c|c|c} \hline \hline
$x\bar{u}_\pi(x)$                 & $(a_0^K)^2$      & $(a_1^K)^2$     & $(a_2^K)^2$     \\ \hline
GRV-P~\cite{grv-p}                & $0.392\pm0.050$ & $0.101\pm0.059$ & $0.507\pm0.077$ \\ 
$2.3\,x^{1.1}(1-x)$~\cite{suecos} & $0.646\pm0.083$ & $0.177\pm0.097$ & $0.177\pm0.127$ \\ 
\hline \hline
\end{tabular}
\end{center}
\end{table}
As can be seen in the Figure, the main effect of using a different $\bar{u}_\pi$ distribution is 
in the low $x$ ($< 0.2$) region. But in this region, the pion valence distribution 
is not well known.

A noticeable effect is also evident in the value of the $a_i^K$ coefficients (see Table~(\ref{table1})). 
In fact, using the GRV-P parametrization, the probability of the $\left|{\overline K}^\circ\pi^-\right>$ 
fluctuation of the kaon is bigger than the probability of the $\left|K_0\right>$ state, 
opposite to intuition. On the other hand, using the $\bar{u}_\pi$ distribution of 
Ref.~\cite{suecos}, one gets $a_1^K\sim a_2^K$ and $a_0^K > a_1^K$, as expected. Given the 
uncertainties coming mainly from the shape of the valon distributions, this facts have no special 
meaning regarding valence distributions in pions. However, the general scheme proposed seems 
to be significant in order to extract kaon's pdf.

\begin{figure}[b] 
\centerline{\psfig{figure=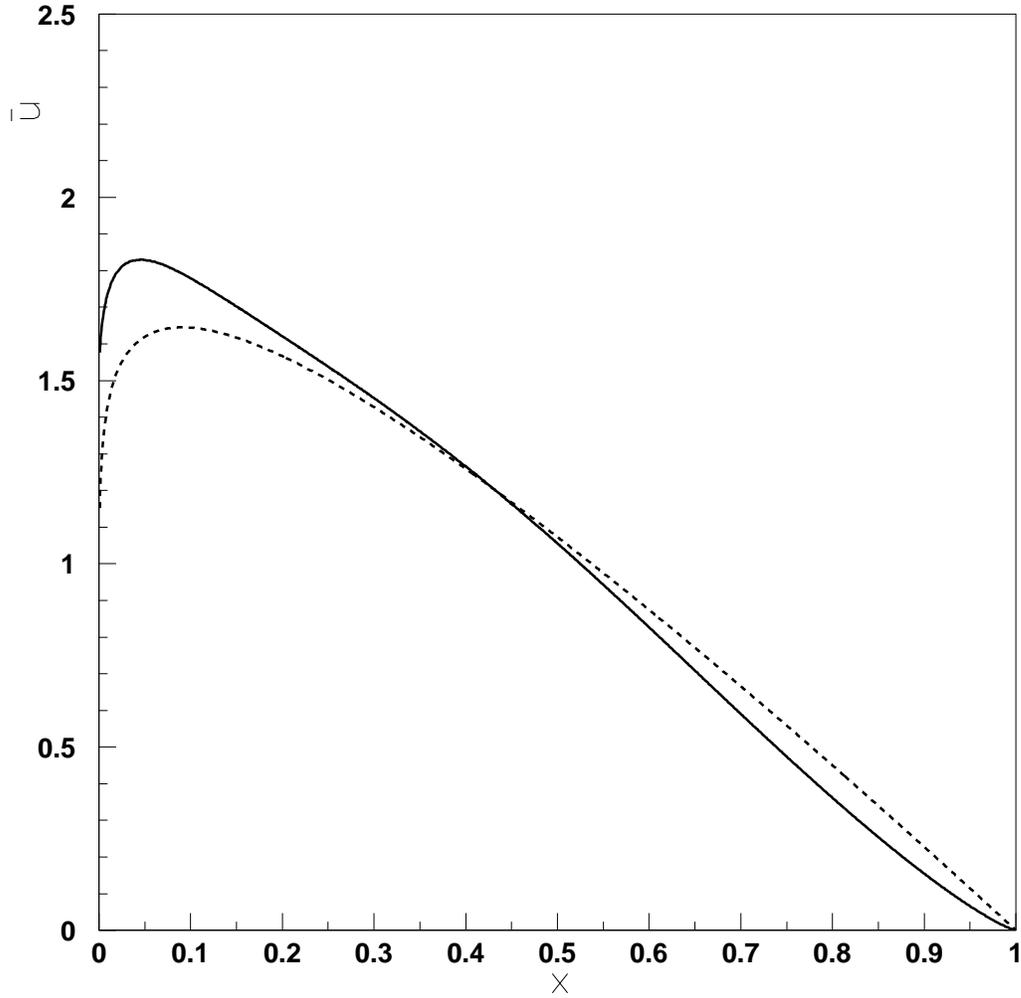,height=6.0in}}
\caption {$\bar{u}_K$ (solid line) compared to the $\bar{u}_\pi$ (dashed
line) distribution as function of x at the low $Q^2_v$ scale. The $\bar{u}_K$ distribution comes from the 
coefficients in the second row of Table~(\ref{table1}). The valence distribution in pions 
uses the the parametrization of Ref.~\cite{suecos}.}
\label{fig2}
\end{figure}

In Fig.~(\ref{fig2}) we display the $\bar{u}_K$ distribution obtained from fits using the $\bar{u}_\pi$ 
parametrization of Ref.~\cite{suecos} in comparison to the $\bar{u}_\pi$ distribution 
itself. As expected, the $\bar{u}_K$ probability density is peaked at lower $x$ than the 
one of the pion, indicating that light valence quarks in kaons carry in average less momentum than 
valence quark in pions.

\begin{figure}[b] 
\centerline{\psfig{figure=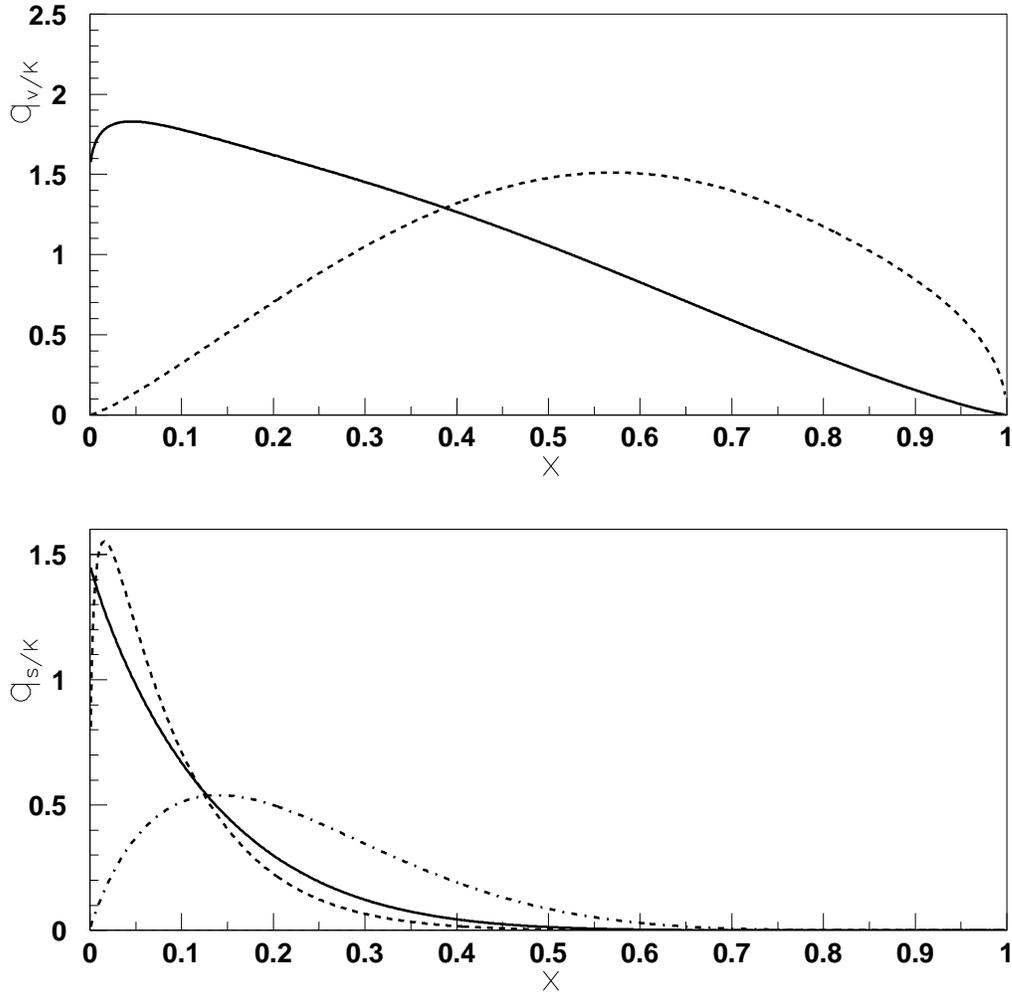,height=6.0in}}
\caption {$K^-$ parton distributions at the $Q_v^2$ scale as a function of $x$ obtained 
from fits using the $\bar{u}_\pi$ parametrization given in Ref.~\cite{suecos}. 
Upper: $\bar{u}_K$ (solid line) and $s_K$ (dashed line) distributions. 
Lower: $d_K$ (solid line), $\bar{d}_K$ (dashed line) and 
$g_K$ (point dashed) distributions.} 
\label{fig3}
\end{figure}

In Fig.~(\ref{fig3}) we show the full set of valence and intrinsic $q\bar{q}$ and gluon 
distributions in kaons at the low $Q^2_v$ scale. These distributions were calculated using 
the coefficients in the second row of Table~(\ref{table1}) together with eqs.~(\ref{eq3}), 
(\ref{eq4}) and (\ref{eq14}).

\section{Conclusions}
\label{sec4}

The study of the low $Q^2$ wave-function of hadrons is an important issue for several reasons. 
The first one is related to the origin of the valence-like distributions needed to fit the 
experimental data on hadron's pdf. This fact, which has already been noted repeatedly by 
several authors in the case of nucleon's pdf, is also important in order to determine the pion 
pdf (see, for instance, the last Ref.~\cite{varios}). Furthermore, for pions there is no 
theoretical input regarding what initial valence-like sea quark or gluon distributions 
are needed in order to describe experimental data. Notice that most of the pion's pdf 
available in the literature are determined using a $SU(3)_{flavor}$ symmetric sea, which is 
surely incorrect. 

The second important reason concerns the kaon structure itself. As a matter of fact, the 
experimental information one can get on the kaon structure is only on the light valence quark 
distribution. The measurement of the strange and even sea quark distributions in kaons 
is not possible due to practical reasons. Actually, strange and sea quarks only contribute 
to the total D-Y dilepton cross section through valence-sea and sea-sea $q\bar{q}$ anihilation. 
Then their contributions are small and cannot be easily separated. Thus, we must rely on other 
methods to obtain information on the kaon structure. It is worth to stress that the knowledge 
of the kaon structure is not only important by itself, but also because kaons are commonly 
used in many experiments and their pdf are needed to calculate the cross sections of the 
produced particles.

In this work we addressed the problem of the low $Q^2$ structure of pions and kaons. We have 
shown, by using a hadron-like Fock state expansion of the pion and kaon wave-functions, 
that the pion and kaon structure are related one another. This enables one to determine the 
complete structure of kaons from a minimal set of measurements of the pion and kaon pdf. 
In fact, by exploiting the relationship among the pion and kaon low $Q^2$ 
wave-functions, the kaon structure can be completely determined by measuring pion structure 
and the $\bar{u}_K/\bar{u}_\pi$ ratio. We would like to remark, however, that a confident 
determination of kaon pdf requires a reanalysis of the pion data in terms of the model presented 
here.

\section*{Acknowledgments}
J.M. is partially supported by COLCIENCIAS, the Colombian Agency for Science 
and Technology, under Contract No. 242-99.

\section*{Appendix: In-meson hadron distributions}

The meson probability density $P_{MM'}(x)$ in the $\left|MM'\right>$ fluctuation has been 
calculated in Refs.~\cite{magnin,christiansen2}. It is given by
\beq
P_{MM'}(x) = \int_0^1{\frac{dy}{y}\int_0^1{\frac{dz}{z} F(y,z)R(x,y,z)}}
\label{eq6}
\eeq
with
\bea
F(y,z) &=& \beta yv_q(y)zq'(z)(1-y-z)^a \nonumber \\
R(x,y,z) &=& \alpha \frac{yz}{x^2}\delta(1-\frac{y+z}{x}) \; .
\label{eq7}
\eea
In eqs.~(\ref{eq7}), $v_q$ and $q'$ are the valon and the quark or antiquark distributions which 
will form the meson $M$ in the $\left|MM'\right>$ fluctuation. The $q'$ distribution is 
generated through the gluon emission from a valon followed by the $q'-\bar{q}\,'$ pair 
creation, which are basic processes in QCD. Thus its distribution is given by~\cite{christiansen}
\beq
q'(x) = \bar{q}'(x) = N \frac{\alpha_{st}^2(Q^2_v)}{(2\pi)^2}
\int_x^1 {\frac{dy}{y} P_{qg}\left(\frac{x}{y}\right) \int_y^1
{\frac{dz}{z} P_{gq}\left(\frac{y}{z}\right) v_q(z)}} \; ,
\label{eq9}
\eeq  
where $P_{qg}(z)$ and $P_{gq}(z)$ are the 
Altarelli-Parisi splitting functions~\cite{a-p} given by
\beq
P_{gq} (z) = \frac{4}{3} \frac{1+(1-z)^2}{z},\ \ \ \ 
P_{qg} (z) = \frac{1}{2} \left( z^2 + (1-z)^2 \right).
\label{eq10}
\eeq

It is worth to note that the only scale dependence appearing in eq.~(\ref{eq9}) arises 
through $\alpha_{st}(Q^2)$. Since the valon scale is tipically of 
the order of $Q_v^2\sim0.64$ GeV$^2$~\cite{hwa}, then the $q'-\bar{q}\,'$ pair creation can 
be safely evaluated perturbatively because $\alpha_{st}^2/(2\pi)^2$ is still sufficiently 
small. The normalization constants $\alpha$, $\beta$ and $N$ in eqs.~(\ref{eq7}) and (\ref{eq9}) 
contribute to the global normalization coefficient of the corresponding Fock state fluctuation in the 
expansions of eqs.~(\ref{eq1}).

Momentum conservation also requires 
\beq
P_{MM'}(x) = P_{M'M}(1-x) \; ,
\label{eq3b}
\eeq
a condition which relates the in-meson $M$ and $M'$ probability densities.
Additionally, hadrons in the $\left|MM'\right>$ fluctuation must be correlated in
velocity in order to form a bound state. This imply that
\beq
\frac{\left<xP_{MM'}(x)\right>}{m_M} = \frac{\left<xP_{M'M}(x)\right>}{m_M'} \; ,
\label{eq3c}
\eeq
fixing in this way the exponent $a$ in eqs.~(\ref{eq7}). Notice also that $P_{gM}$ 
is calculated from the ``recombination'' of an antiquark with a valon of the same 
flavor~\cite{christiansen2}. Then, formally, the $P_{gM}$ corresponds to the 
$\pi^\circ$ distribution in a hypothetical $\left|M\pi^\circ\right>$ fluctuation.

\newpage

\end{document}